\documentclass[a4paper,12pt]{article}
\usepackage[cp1251]{inputenc}
\usepackage[russian]{babel}

\usepackage{amsmath}

\def\bb{\begin{equation}}
\def\ee{\end{equation}}

\begin{document}

\vskip 0.4cm
\title{ `` Quantizations''  of higher hamiltonian analogues of the Painleve I and Painleve II equations with two degrees of freedom}
\author{ B. I. Suleimanov{\footnote{e-mail: bisul@mail.ru}}} 
\date{}
\maketitle

\begin{abstract}
We construct a solution of an analog of the Schr\"{o}dinger equation for the Hamiltonian $ H_1 (z, t, q_1, q_2, p_1, p_2) $ corresponding to the second equation $P_1^2$ in the Painleve I hierarchy. This solution is produced by an explicit change of variables from a solution of the linear equations whose compatibility condition  is the ordinary differential equation $P_1^2$ with respect to $z$. This  solution also satisfies an analog of the Schr\"{o}dinger equation corresponding to the Hamiltonian $ H_2 (z, t, q_1, q_2, p_1, p_2) $ of Hamiltonian system with respect to $t$ which is compatible with $P_1^2$. A similar situation  occurs for the $P_2^2$ equation in the Painleve II hierarchy.
\end{abstract}

\newpage

\vskip 0.4cm
\begin{center}
\Large{ ``Квантования''  высших гамильтоновых аналогов уравнений Пенлеве I и II с двумя степенями свободы}

\large{ Б. И. Сулейманов} 
\end{center}

\begin{abstract}
Построено решение аналога уравнения Шредингера, определяемого гамильтонианом $H_1(z,t,q_1, q_2, p_1, p_2)$ второго члена $P_1^2$ иерархии первого уравнения Пенлеве. После явной замены оно   задается решением систем линейных уравнений, 
условием совместности которых является  нелинейное обыкновенное дифференциальное уравнение $P_1^2$ по независимой переменной $z$. Результат этой   замены удовлетворяет также аналогу уравнения Шредингера, определяемого гамильтонианом $H_2(z,t,q_1, q_2, p_1, p_2)$  гамильтоновой системы c независимой переменной $t$, которая совместна с уравнением $P_1^2$. Показано, что схожая ситуация  имеет место для представителя $P_2^2$ иерархии второго уравнения Пенлеве.
\end{abstract}

Для всех шести  обыкновенных дифференциальных уравнений (ОДУ) Пенлеве $q_{zz}=f(z,q, q_z)$, решения соответствующих им 
пар линейных систем  метода изомонодромных деформаций (ИДМ) из статьи Р.Гарнье \cite{Garn}  с помощью явной замены задают \cite{Difufa} решения  уравнений \begin{equation}\label{murze}\frac{\partial}{\partial z}\Psi=H(z,x,-\frac{\partial}{\partial x})\Psi\end{equation}
(см. также начало Заключения данной статьи). Эти уравнения определяются квадратичными по импульсам $p$ гамильтонианами $H=H(z,q,p)$ гамильтоновых систем 
$q'_z=H'_p(z,q,p),$ $p'_z=-H'_q(z,q,p)$,
исключение из которых $p$ дает шесть ОДУ Пенлеве.  
Из  уравнений Шредингера
\begin{equation}\label{Schroo}\varepsilon \frac{\partial}{\partial z}\Psi=
H(z,x,-\varepsilon\frac{\partial}{\partial x})\Psi,\end{equation} зависящих от постоянной Планка $h=2\pi\hbar=-2\pi i\varepsilon$, эти шесть эволюционных уравнений (\ref{murze})  получаются в результате формальной замены $\varepsilon=1$. (Для случаев уравнений Пенлеве III и V в формулах \cite{Difufa} содержатся неточности. Но они легко поправимы.)

Отметим, что для ОДУ, получающихся из первого и второго уравнений Пенлеве I и II, 
соответствующие уравнения ИДМ после растяжений  переходят в cовместные системы линейных ОДУ, которые при произвольном фиксированном комплексном числе $\varepsilon$ задают точные решения уравнений (\ref{Schroo}), определяемые гамильтонианами $H$, зависящими от $\varepsilon$. Поясним сказанное:

 Пара уравнений ИДМ Гарнье \cite{Garn} для ОДУ 
($a_j$ --- произвольные постоянные)
 \begin{equation}\label{peodg}\lambda_{\tau\tau}=a_4(2\lambda^3+\tau \lambda)+a_3(6\lambda^2+\tau)+a_2\lambda+a_1,\end{equation}
 которое в частных случаях содержит первое и второе уравнения Пенлеве, имеет вид
\begin{equation} \label{garlpeodg}W_{yy}=P(\tau,y)W,\qquad W_{\tau}=B(\tau,y)W_y-\frac{1}{2}B_y(\tau,y)W,\end{equation} 
где 
$$B=\frac{1}{2(y-\lambda)}, \quad P=a_4[y^4-\lambda^4+\tau(y^2-\lambda^2)]+2a_3[2(y^3-\lambda^3)+\tau(y-\lambda)]+$$
$$+a_2(y^2-\lambda^2)+2a_1(y-\lambda)+(\lambda')^2-\frac{\lambda'}{y-\lambda}+\frac{3}{4(y-\lambda)^2}.$$
Уравнения (\ref{garlpeodg}) с такими коэффициентами $B(\tau,y)$ и $P(\tau,y)$ совместны на решениях $\lambda(\tau)$ ОДУ (\ref{peodg}). Замена 
$V=\sqrt{(y-\lambda)}W$ систему (\ref{garlpeodg}) переводит в  уравнения 
\begin{equation} \label{rashpeod}V_{yy}=\frac{V_y}{y-\lambda}+[P-\frac{3}{4(y-\lambda)^2}]V,\qquad V_{\tau}=\frac{V_y-\lambda'V}{2(y-\lambda)},\end{equation} 
Из них следут, что для их совместного решения $V(\tau,y)$ справедливо тождество
\begin{equation}\label{kkpeod}V_{\tau}=\frac{V_{yy}}{2}-[\frac{a_4}{2}(y^4+\tau y^2)+a_3(2y^3+\tau y)+
\frac{a_2}{2}y^2+a_1z+H(\tau,\lambda(\tau),\lambda'(\tau)]V.\end{equation} 
Здесь  функция $H(\tau,\lambda(\tau),\lambda'(\tau))$
при $\lambda=\lambda(\tau)$ и 
$\mu=\lambda'(\tau)$ совпадает с гамильтонианом 
\begin{equation}\label{hamot}H=\frac{\mu^2}{2}-\frac{a_4}{2}(\lambda^4+\tau \lambda^2)-a_3(2\lambda^3+\tau \lambda)-\frac{a_2}{2}\lambda^2-a_1\lambda\end{equation}
гамильтоновой системы 
$\lambda'_{\tau}=H'_{\mu}(\tau,\lambda,\mu),$ $\mu'_{\tau}=-H'_{\lambda}(\tau,\lambda,\mu),$ 
эквивалентной ОДУ (\ref{peodg}). Преобразованием 
$\Psi=\exp{(\int_{\tau_*}^{\tau}H(\nu,\lambda(\nu),\mu(\nu))d\nu)}V $
тождество (\ref{kkpeod}) сводится к эволюционному уравнению вида (\ref{murze}) 
$$\Psi_{\tau}=\frac{\Psi_{yy}}{2}-[\frac{a_4}{2}(y^4+\tau y^2)+a_3(2y^3+\tau y)+\frac{a_2}{2}y^2+a_1y]\Psi,$$
определяемому гамильтонианом (\ref{hamot}). Последнее же в результате растяжений 
$$z=\varepsilon \tau,\quad x=\varepsilon y,\quad q=\varepsilon \lambda  $$
переходит в уравнение вида (\ref{Schroo})
$$\varepsilon\Psi_{z}=\frac{\varepsilon^2\Psi_{xx}}{2}-[\frac{a_4}{2}(\frac{x^4}{\varepsilon^4}+\frac{z x^2}{\varepsilon^3})+a_3(\frac{2x^3}{\varepsilon^3}+
\frac{zx}{\varepsilon^2})+\frac{a_2}{2}\frac{x^2}{\varepsilon^2}+a_1\frac{x}{\varepsilon}]\Psi,$$
определяемого гамильтонианом   
$$H(z,q,p)=\frac{p^2}{2}-\frac{a_4}{2}(\frac{q^4}{\varepsilon^4}-\frac{z q^2}{\varepsilon^3})-a_3(\frac{2q^3}{\varepsilon^3}-
\frac{zq}{\varepsilon^2})-\frac{a_2}{2}\frac{q^2}{\varepsilon^2}-a_1\frac{q}{\varepsilon}.$$

Следуя терминологии \cite{Tmf}, уравнения типа уравнений Шредингера (\ref{Schroo}) с постоянными $\varepsilon$, не связанными с  $h$,  называются в этой статье ``квантованиями''. ( При $\varepsilon=1$ они используются, например,  при описании фильтрации процессов диффузии  \cite{Ovs}, \cite{Dovs}.)

Дальнейшее развитие результаты \cite{Difufa} получили в работах \cite{Tmf}, \cite{Novd}---\cite{Ufmj}. Но вопрос о подобных ``квантованиях'' для  высших аналогов уравнений Пенлеве до сих пор не изучался. Ниже такие ``квантования'' приводятся для  совместных решений уравнений Кортевега --- де Вриза (КдВ) 
\begin{equation}\label{KdV}u_t=-uu_z-u_{zzz}\end{equation}
и Нелинейного уравнения Шредингера (НУШ) 
\begin{equation}\label{nnushv}
-iu_t=u_{zz}+2\delta|u|^2u \qquad (\delta=const\in R)
\end{equation}
с ОДУ, определяемых суммами стационарных частей  первых высших автономной симметрий уравнений (\ref{KdV}), (\ref{nnushv}) и их симметрий Галилея.  Эти высшие аналоги, соответственно, ОДУ Пенлеве I и II  эквивалентны двум парам совместных гамильтоновых систем с двумя степенями свободы по независимым переменным $z$ и $t$.

\section{Член $P_1^2$ иерархии уравнения Пенлеве I}

{\bf 1.1.} Первым высшим представителем $P_1^2$ иерархии изомонодромных нелинейных ОДУ $ P_1^n$  уравнения Пенлеве I,  называемых также  массивными $(2n+1,2)$ струнными уравнениями
\cite{Mo}, \cite{Kit}, является уравнение
\begin{equation}u_{zzzz}+\frac{5}{3}uu_{zz}+
\frac{5}{6}u^2_z+\frac{5}{18}(z-tu+u^3)=0,\qquad t=const.\label{GP}\end{equation}

Интерес к ОДУ (\ref{GP}) в первую очередь связан с тем, что оно обладает специальным решением $u(z,t)$, возникающим при исследовании самых разных задач математической физики. В частности \cite{s120},  это решение cовпадает с решением Гуревича---Питаевского (ГП) 
уравнения КдВ (\ref{KdV}), введенным в рассмотрение в \cite{g77} в качестве функции, универсальным образом описывающая влияние малых дисперсионных добавок на опрокидывание простых волн в нелинейной гидродинамике. (В главном порядке асимптотика  специального решения ГП при  $t\to \infty $ описана в известной работе \cite{g78}.  Результаты \cite{g78} уточнены в \cite{s120}, \cite{k62}---\cite{esaul}, в  \cite{ClV} 
доказана его гладкость, в \cite{gst} поведение решения ГП промоделировано численно.) Подобную же роль эта специальная функция играет не только в случае простых волн, но и \cite{k62}, \cite{k63}, \cite {kub}---\cite{Dubl}  для решений общего положения двухкомпонентных гидродинамических систем с малой дисперсией. Ряд конкретных   двухкомпонентных систем, большей частью не интегрируемых, решения которых подтверждают справедливость  последнего утверждения, описан в  \cite{k63}--- см. также \cite{gst}, \cite{dumn}, \cite{Dubl}. А согласно гипотезе, сформулированной \linebreak Б.  А. Дубровиным  \cite{dumn}, \cite{Dubl}, решение ГП для задач с малой дисперсией должно играть еще более универсальную роль. Кроме того \cite{s120}, \cite{gst}, эта же нелинейная специальная функция  изучалась  в \cite{bresm}, \cite{dous} в связи с задачами квантовой теории гравитации. 
 
{\bf 1.2.} Совместные решения уравнений (\ref{KdV}) и (\ref{GP}) относятся к  классу изомононодромных решений уравнений нулевой кривизны \cite{itiz}. Эти уравнения есть  условие совместности линейных систем ( всюду далее нижний индекс при $u$ означает порядок производной $u$ по переменной $z$) 
\begin{equation}\label{mozkx}
\Phi_z=\left(\begin{array}{cc}0 &1\\\zeta-u/6&0\end{array}\right)\Phi,
\Phi_t=\left(\begin{array}{cc}u_1/6&-u/3-4\zeta\\ u_2/6+u^2/18+\zeta u/3-4\zeta^2&-u_1/6\end{array}\right)\Phi,
\end{equation}

\begin{equation}\label{isok}
\frac{5}{1728}\Phi_{\zeta}=(\frac{4\zeta u_1+u_3+uu_1}{96}\left(\begin{array}{cc}-1&0\\0&1\end{array}\right)
+(\zeta^2+\frac {\zeta u}{12}+\frac{u_2}{48}+\frac{u^2}{96}-\frac{5t}{288}) \left(\begin{array}{cc}0 &1\\0&0\end{array}\right)+$$
$$+(\zeta^3-\frac{\zeta^2}{12}-\zeta\frac{6u_2+u^2+5t}{288}+\frac{u_2u}{288}-\frac{u_1^2}{576}+\frac{u^3}{864}+\frac{5z}{1728})\left(\begin{array}{cc}0 &0\\1&0\end{array}\right))\Phi.
\end{equation}

\mbox{З а м е ч а н и е 1.}  Системы  (\ref{mozkx}), (\ref{isok}), взятые из статьи \cite{ClV},  простыми заменами сводятся к 
трем линейных системам, выписанным ранее в \cite{s120}.

При этом имеет место 
система  ОДУ, состоящая из  (\ref{KdV}) и уравнений  
$$(u_1)_t=\frac{2}{3}uu_2-\frac{1}{6}u_1^2+\frac{5}{18}(z-tu+u^3),$$
\begin{equation}\label{GPKV}(u_2)_t=\frac{2}{3}uu_3+\frac{1}{3}u_1u_2+\frac{5}{18}(1-tu_1+3u_1u^2),\end{equation}
$$(u_3)_t=u_3u_1+\frac{1}{3}(u_2)^2-\frac{5}{18}u^2u_2+\frac{10}{9}u(u_1)^2-\frac{5}{27}(u^4-tu^2+zu)
-\frac{5}{18}tu_2.$$

ОДУ (\ref{GP}) и данная система ОДУ эквивалентны гамильтоновым системам с двумя степенями свободы --- соответственно, системам
\begin{equation} \label{Hamsyso}(q_j)'_z=(H_1)'_{p_j}, \qquad (p_j)'_z=-(H_1)'_{q_j}\qquad(j=1,2),\end{equation}
\begin{equation} \label{Hamsyst}(q_j)'_t=(H_2)'_{p_j}, \qquad (p_j)'_t=-(H_2)'_{q_j}\qquad(j=1,2),\end{equation}
c квадратичными по импульсам $p_j$ гамильтонианами 
\begin{equation} \label{Hamone}H_1(z,t,q_1,q_2,p_1,p_2)=p_1p_2 -\frac{5q_1^4}{8}+\frac{5q_2q_1^2}{2}-\frac{q_2^2}{2}-\frac{5tq_1^2}{36}+\frac{5zq_1}{108},\end{equation}
\begin{equation} \label{Hamtwo}H_2(z,t,q_1,q_2,p_1,p_2)=-\frac{p_1^2}{2}-q_1p_1p_2+
(q_2-\frac{5t}{36})p_2^2+\frac{5p_2}{108}+$$
$$+\frac{q_1^5}{2}-2q_1q_2^2-\frac{5tq_1^3}{36}+\frac{5tq_1q_2}{18}-\frac{5zq_1^2}{216}-\frac{5zq_2}{108},\end{equation}
где
$q_1=u/6$, $q_2=u_2/6+5u^2/72$ 
$p_1=u_3/6+5uu_1/36$, $p_2=u_1/6$
(эти формулы получены, исходя из аналогий с формулами (5.1.7) --- (5.1.10) статьи \cite{Dubun} для двухзонных решений уравнения КдВ). Ниже показано, что каждое решение (\ref{mozkx}), (\ref{isok})  задает решение уравнений ($\varepsilon=5/54$)
\begin{equation}\label{qnGpz}\varepsilon\Psi_{z}=\varepsilon^2\Psi_{xy}+
[-\frac{5x^4}{8}+\frac{5x^2y}{2}-\frac{y^2}{2}
-\frac{5tx^2}{36}+\frac{5zx}{108}]\Psi, \end{equation}
$$\varepsilon \Psi_{t}=\varepsilon^2[
-\frac{\Psi_{xx}}{2}-x \Psi_{xy}
+\frac{1}{2}\frac{\partial}{\partial y}((y-\frac{5t}{36}) \Psi_{y})+(y-\frac{5t}{36})\frac{\Psi_{yy}}{2}]
-\varepsilon\frac{5\Psi_y}{108}+$$
\begin{equation}\label{qnGpt}+[\frac{x^5}{2}-2xy^2
-\frac{5tx^3}{36}+\frac{5txy}{18}-\frac{5zx^2}{216}-\frac{5zy}{108}]\Psi, \end{equation}
являющихся ``квантованиями'', определяемыми гамильтонианами (\ref{Hamone}) и (\ref{Hamtwo}) --- после формальных замен
$q_1\to x,$ $q_2\to y$  и $p_1\to -\varepsilon \frac{ \partial}{\partial x}$, $p_2\to -\varepsilon \frac{ \partial}{\partial y} $ их
 можно символически записать в  виде: 
\begin{equation}\label{QHamst}\varepsilon\Psi _z=H_1(z,t,,x,y,-\varepsilon \frac{\partial}{\partial x},-\varepsilon  \frac{\partial}{\partial y})\Psi,\quad \varepsilon\Psi _t=H_2(z,t,x,y,-\varepsilon\frac{\partial}{\partial x},-\varepsilon \frac{\partial}{\partial y})\Psi.\end{equation}

 {\bf 1.3.} Фундаментальные решения ОДУ (\ref{mozkx}), (\ref{isok}) задают $2\times2$ 
  матрицы 
\begin{equation}\label{NovDP}M(z,t,\eta,\zeta)=
\Phi^{-1}(z,t,\eta)\Phi(z,t,\zeta),\end{equation} 
которые удовлетворяют двум скалярным линейным уравнениям 
$$\frac{1728}{5}(\zeta-\eta) M_{z}=M_{\zeta\zeta}-M_{\eta\eta}-2\frac{M_{\zeta}+M_{\eta}}{\zeta-\eta}-$$
\begin{equation}\label{NGPez}-(\frac{1728}{5})^2[(\zeta^5-
\eta^5)-\frac{5t(\zeta^3-\eta^3)}{144}+\frac{5z(\zeta^2-\eta^2)}{1728}+r_1(t,z)(\zeta-\eta)]M, \end{equation}
$$\frac{1728}{20}(\zeta-\eta) M_{t}=\eta M_{\zeta\zeta}-\zeta M_{\eta\eta}-(\zeta+\eta)\frac{M_{\zeta}+M_{\eta}}{\zeta-\eta}+$$
\begin{equation}\label{NGPet}+(\frac{1728}{5})^2[\zeta(\eta^5-\frac{5t\eta^3}{144}+\frac{5z\eta^2}{1728}+r_0(t,z))-
\eta(\zeta^5-\frac{5t\zeta^3}{144}+\frac{5z\zeta^2}{1728}+r_0(t,z))]M,
 \end{equation}
зависимость которых от $u(z,t)$ содержится лишь в коэффициентах 
$$r_1(z,t)=\frac{1}{48^2}(2u_3u_1-u_2^2+\frac{5u_1^2u}{3}+\frac{5zu}{9}+\frac{5u^4}{36}-\frac{5tu^2}{18}+\frac{25t^2}{36}),$$ 
$$r_0(z,t)=\frac{1}{96^2}(u_3^2+2u_3u_1u+\frac{2u_2^2u}{3}-
\frac{u_2u_1^2}{3}+\frac{5u_2u^3}{3}+\frac{5u_2(z-tu)}{9}+
\frac{5u_1^2u^2}{6}+$$
$$+\frac{5tu_1^2}{36}+\frac{u^5}{9}-\frac{5tu^3}{27}+\frac{5zu^2}{18}-
\frac{25zt}{54}). $$ 

Заменой
\begin{equation}\label{chGP}M=(\zeta-\eta)\exp{S(t,z)}W,\end{equation}
где функция $S$ удовлетворяет непротиворечивым равенствам
$$S_z=-\frac{1728}{5}[r_1(t,z)-\frac{25t^2}{(288)^2}],\qquad S_t=\frac{1728}{5}[4r_0(t,z)+\frac{50zt}{(288)^2}],$$
уравнения (\ref{NGPez}), (\ref{NGPet}) переводятся в независящие от $u(z,t)$ уравнения  
\begin{equation}\label{KGPez} \frac{1728}{5} W_{z}=\frac{W_{\zeta\zeta}-W_{\eta\eta}}{\zeta-\eta}-$$
$$-
(\frac{1728}{5})^2[\frac{\zeta^5-
\eta^5}{\zeta-\eta}-\frac{5t(\zeta^3-\eta^3)}{144(\zeta-\eta)}+\frac{5z(\zeta^2-\eta^2)}{1728(\zeta-\eta)}+
\frac{25t^2}{(288)^2}]W, 
\end{equation}
$$\frac{1728}{20} W_{t}=\frac{\eta W_{\zeta\zeta}-\zeta W_{\eta \eta}}{\zeta-\eta}-\frac{W_{\zeta}-W_{\eta}}{\zeta-\eta}
+(\frac{1728}{5})^2[\zeta(\eta^5-
\frac{5t\eta^3}{144}+\frac{5z\eta^2}{1728}-$$
\begin{equation}\label{KGPet}-\frac{25zt}{ (288)^26})
-\eta(\zeta^5-\frac{5t\zeta^3}{144}+\frac{5z\zeta^2}{1728}+\frac{25zt}{ (288)^26})]\frac{W}{\zeta-\eta}. \end{equation}
После перехода от $\zeta$ и $\eta$ к независимым переменным 
\begin{equation}x=-\frac{\zeta+\eta}{2},\quad y=-\frac{(\zeta-\eta)^2}{2}+\frac{5t}{144}\label{chanew}\end{equation}
уравнения (\ref{KGPez}), (\ref{KGPet}) принимают вид ``квантований'' (\ref{qnGpz}) и (\ref{qnGpt}).

\section{Член $P_{2}^2$ иерархии уравнения Пенлеве II}

{\bf 2.1.}  Для вещественных значений $z$ и $t$ ниже описываются ``квантования'' совместных решений  НУШ (\ref{nnushv}) и ОДУ 
\begin{equation}\label{nnushxv}
\beta u_3-4tu_1+6\beta \delta|u|^2u_1+2iz u=0.
\end{equation}
Наряду с решением ГП уравнения КдВ, такие решения также играют важную роль для задач математической физики с малым параметром:

---  они задают \cite{lombis}, \cite{kitj} специальные решения Хабермана --- Сана НУШ (\ref{nnushv}) из работы \cite{h93}, которые  в пределе при $\delta \to 0$ переходят  
в интеграл Пирси \linebreak $Q  = const\int \limits_{-\infty}^{\infty} \exp[-2i(\beta \lambda^4 +2t\lambda^2+ x\lambda)]d\lambda$, и которые для довольно широкого ряда cитуаций  описывают влияние малых нелинейностей на высокочастотные асимптотики около острия (клюва) каустики;
 
---  при $\delta>0$  с помощью других решений уравнений (\ref{nnushv}) и
(\ref{nnushxv}) описывается \cite{kuph} влияние малой дисперсии на процессы провального самообострения импульса, которые характерны для приближений нелинейной геометрической оптики  с нелинейностью общего положения;

---  вообще, ОДУ (\ref{nnushxv}) есть первый высший представитель $P_2^2$  иерархии $ P_2^n$  уравнения Пенлеве II,  важность роли которой (наряду с важностью роли иерархии $P_1^n$) для широкого класса задач с малым параметром  была предсказана А. В. Китаевым в \cite{Kit}, исходя из аналогий с иерархиями интегралов Фурье канонического вида, называемых \cite[Гл. VI, \S4]{Fed} специальными функциями волновых катастроф. По поводу аналогий  с уравнением Пенлеве $II$ после просмотра раздела 2.4 данной статьи cм. еще раздел 6.1 работы К. Окамото \cite{Okom}.

{\bf 2.2.} Совместные решения уравнений (\ref{nnushv}) и (\ref{nnushxv}) также изомонодромны. Они есть \cite{lombis} условие совместности трех линейных уравнений 
\begin{equation}\label{lnush}
\Phi_{z}=i
\left(\begin{array}{cc}-\zeta&u\\\delta u^*&\zeta\end{array}\right)\Phi, \quad 
\Phi_{t}=
\left(\begin{array}{cc}-i(2\zeta^2-\delta |u|^2) &2i\zeta u-u_1\\
\delta(2i\zeta u^*+u^*_1)&i(2\zeta^2-\delta |u|^2)\end{array}\right)\Phi,
\end{equation}
$$\Phi_{\zeta}=(-[i(4\beta \zeta^3+4\zeta t-2\delta \beta|u|^2\zeta+z)+
\delta \beta(u_1u^*-u^*_1u)]\left(\begin{array}{cc}1&0\\0&-1\end{array}\right)+$$
\begin{equation}\label{isomgsh}
+\left(\begin{array}{cc}0&4i\beta\zeta^2u-2\beta\zeta u_1+4itu-\beta u_t\\ \delta (4i\beta\zeta^2u^*+2\beta\zeta u_1^*+4itu^*+\beta u^*_t)&0\end{array}\right))\Phi,
\end{equation}
где $u^*$ --- комплексное сопряжение $u$. При этом  $u$ задает \cite{lombis} решения системы ОДУ по переменной $t$, определяемой НУШ (\ref{nnushv}) и уравнениями $$\beta u_{tt}=4itu_t+2i\beta \delta |u|^2u_t+(2i+8t \delta |u|^2)u+
2izu_1+2\beta \delta u_1(u_1u^*-u_1^*u),$$
\begin{equation}\label{nusht}
\beta(u_1)_t=2zu+4itu_1-2i\beta \delta u(u_1u^*-u_1^*u).
\end{equation}
Из уравнений (\ref{nnushv}), (\ref{nnushxv})и (\ref{nusht}) cледует постоянство по $z$ и  $t$ выражения
\begin{equation}\label{konsh}\frac{c}{4\delta \beta^2}=-u_2^*u-u_2u^*+|u_1|^2-3\delta|u|^4+\frac{4t}{\beta}|u|^2=
i(u_tu^*-u_t^*u)+|u_1|^2+\delta|u|^4+\frac{4t}{\beta}|u|^2.\end{equation}

{\bf 2.3.} Формулой (\ref{NovDP}) фундаментальные решения ОДУ (\ref{lnush}), (\ref{isomgsh}) определяют $2\times2$ 
 матрицы $M(t,z,\zeta,\eta)$, которые удовлетворяют уравнениям 
 $$4\beta(\zeta-\eta) M_{z}-M_{\zeta\zeta}+M_{\eta\eta}+2\frac{M_{\zeta}+M_{\eta}}{\zeta-\eta}=[16\beta^2(\zeta^6-\eta^6)+$$
\begin{equation}\label{Nshez}+32\beta t(\zeta^4-\eta^4)+8\beta z(\zeta^3-\eta^3)+(16t^2+c)(\zeta-\eta)^2+r_1(z,t)(\zeta-\eta)]M, \end{equation}
$$2\beta(\zeta-\eta)M_{t}-\eta M_{\zeta \zeta}+\zeta M_{\eta\eta}+\frac{(\zeta+\eta)(M_{\zeta}+M_{\eta})}{\zeta-\eta}=[\zeta(16\beta^2\eta^6+32\beta t\eta^4+8\beta z\eta^3+$$
\begin{equation}\label{Nshet} +(16t^2+c)\eta^2+r_0(z,t))-\eta(16 \beta^2 \zeta^6+32\beta t\zeta^4+8\beta z\zeta^3+(16t^2+c)\zeta^2+r_0(z,t))]M,
 \end{equation}
где $c$ --- таже постоянная, что в формуле (\ref{konsh}),  
$$r_1(z,t)=2i\delta \beta^2(u_2u_1^*-u_2^*u_1)-4\delta z|u|^2+8zt,$$
$$r_0(z,t)=16t^2\delta|u|^2-16\delta^2 \beta t|u|^4-4\delta\beta t(u_2^*u+u_2u^*)+\delta \beta^2|u_2|^2+
4\delta^3\beta^2|u|^6+$$
$$+2\delta^2\beta^2|u|^2(u_2^*u+u_2u^*)-\delta^2\beta^2(u_1u^*-u_1^*u)^2-2i\delta \beta z(u_1u^*-u_1^*u)+z^2$$ 
Функции $r_1$ и $r_2$ связаны соотношением  $(r_1(z,t))'_t=2(r_0(z,t))'_z+4z.$

Заменой (\ref{chGP}),  
где функция $S$ такова, что
$$4\beta S_z=r_1(z,t),\quad 2\beta S_t=r_0(t,z)+ z^2,$$
уравнения (\ref{Nshez}), (\ref{Nshet}) переводятся в пару уравнений  
$$4\beta W_{z}-\frac{W_{\zeta\zeta}-W_{\eta\eta}}{\zeta-\eta}=$$
\begin{equation}\label{KNez}=[16\beta^2(\zeta^6-
\eta^6)+32\beta t(\zeta^4-\eta^4)+8\beta z(\zeta^3-\eta^3)+
(16t^2+c)(\zeta^2-\eta^2)]\frac{W}{\zeta-\eta}, \end{equation}
$$2\beta W_{t}-\frac{\zeta W_{\eta\eta}-\eta W_{\zeta \zeta}}{\zeta-\eta}-\frac{W_{\zeta}-W_{\eta}}{\zeta-\eta}
=[\zeta(16\beta^2\eta^6+32\beta t\eta^4+8\beta z\eta^3+(16t^2+c)\eta^2-$$
\begin{equation}\label{KNet}
-z^2)-\eta(16\beta^2\zeta^6+32\beta t\zeta^4+8\beta z\zeta^3+(16t^2+c)\zeta^2-z^2)]\frac{W}{\zeta-\eta},  \end{equation}
не содержащих зависимости от $u$.

{\bf 2.4.} H. Kimura в \cite{kimu} привел список совместных пар гамильтоновых изомонодромных систем с двумя степенями свободы, получающийся в результате 
последовательных вырождений системы Р.Гарнье из статьи \cite{Garn}. В их числе содержится пара, нумеруемая в \cite{kimu} как $H(5)$:
\begin{equation} \label{Kamsyso}
(\lambda_j)'_{t_1}=(K_1)'_{\mu_j}, \quad (\mu_j)'_{t_1}=-(K_1)'_{\lambda_j}\qquad(j=1,2),
\end{equation}
\begin{equation} \label{Kamsyst}
(\lambda_j)'_{t_2}=(K_2)'_{p_j},\quad (\mu_j)'_{t_2}=-(K_2)'_{\lambda_j}\qquad(j=1,2),\end{equation}
c квадратичными по импульсам $\mu_j$ гамильтонианами 
\begin{equation} \label{Kamone}K_1(t_1,t_2,\lambda_1,\lambda_2,\mu_1,\mu_2)=
\frac{1}{2}\sum_{k=1}^2\frac{1}{\Lambda'(\lambda_k)}[\mu_k^2-P(t_1,t_2,\lambda_k)\mu_k-2\nu\lambda_k^2],\end{equation}
\begin{equation} \label{Kamtwo}K_2(t_1,t_2,\lambda_1,\lambda_2,\mu_1,\mu_2)=\frac{1}{2}
\sum_{k=1}^2\frac{Q(\lambda_k)}{\Lambda'(\lambda_k)}[\mu_k^2-(P(t_1,t_2,\lambda_k)+
\frac{1}{Q(\lambda_k)})\mu_k-2\nu\lambda_k^2],
\end{equation}
где $\nu$ --- постоянная, $P(t_1,t_2,q)=2q^3+2t_2q+t_1$, $\Lambda(q)=(q-\lambda_1)(q-\
\lambda_2
)$, $\Lambda'(q)$ --- производная $\Lambda(q)$ по $q$,  $Q(q)=q-\lambda_1-\lambda_2$. (В \cite{kimu} имеется опечатка, повторенная в \cite{Okom}:  гамильтониан $K_2$ приведен без множителя {1/2}.)

ОДУ (\ref{nnushxv}) по переменной $z$  и система ОДУ по переменной $t$, определяемой НУШ (\ref{nnushv}) и уравнениями  (\ref{nusht}) после замен \begin{equation}\label{channtwo}z=\frac{i\alpha t_1}{2} , \quad t=\frac{\beta t_2}{ \alpha}, \quad \alpha=(4 \beta)^{1/4}\exp{(i\pi/8)},\end{equation} эквивалентны гамильтоновым системам (\ref{Kamsyso}) и, cоответственно, (\ref{Kamsyst}),  определяемых   гамильтонианами (\ref{Kamone}) и (\ref{Kamtwo}).  При этом 
$$\lambda_1=\alpha(\frac{iu_z}{4u}-\sqrt{-\frac{(u_z)^2}{16u^2}-
\frac{t}{ \beta}+\frac{u_{zz}}{4u}+\frac{\delta |u|^2}{2}}),$$ 
$$\lambda_2=\alpha(\frac{iu_z}{4u}+\sqrt{-\frac{(u_z)^2}{16u^2}-
\frac{t}{ \beta}+\frac{u_{zz}}{4u}-\frac{\delta |u|^2}{2}}),$$ 
$$\mu_1=\frac{\delta \alpha^3}{2}(\frac{i}{4}(2u_zu^*-u^*_zu)-|u|^2\sqrt{-\frac{(u_z)^2}{16u^2}-\frac{t}{\beta}+\frac{u_{zz}}{4u}+\frac{\delta|u|^2}{2}}),$$
$$\mu_2=\frac{\delta \alpha^3}{2}(\frac{i}{4}(2u_zu^*-u^*_zu)+|u|^2\sqrt{-\frac{(u_z)^2}{16u^2}-\frac{t}{\beta}+\frac{u_{zz}}{4u}+\frac{\delta|u|^2}{2}}),$$
где постоянная $\alpha$  определена заменами (\ref{channtwo}), $\nu=ic/(8\beta)$. (Последние  формулы были получены автором, исходя из уравнений ИДМ (\ref{lnush}), (\ref{isomgsh}) и формул (6.3), (6.4) работы \cite{Okom}---cм. также \cite{kimu}. )

{\bf 2.5.} Совместна следующая пара эволюционных уравнений 
$$2(X-Y)\Gamma_{t_1}=\Gamma_{XX}-\Gamma_{YY}+2X^2(X\Gamma)_{X}
+(2t_2X+t_1)\Gamma_{X}-2Y^2(Y\Gamma)_Y-$$
\begin{equation}\label{Kamz}-(2t_2Y+t_1)\Gamma_{Y}
-2\nu(X^2-Y^2)\Gamma,\end{equation}
$$2(X-Y)\Gamma_{t_2}=X[\Gamma_{YY}+2Y^2(Y\Gamma)_{Y}+(2t_2Y+t_1)\Gamma_{Y}-2\nu Y^2\Gamma]-$$
\begin{equation}\label{Kamt}-Y[\Gamma_{XX}+(2X^2)(X\Gamma)_{X}+(2t_2X+t_1)\Gamma_{X}-2\nu X^2\Gamma]+\Gamma_X-\Gamma_Y,\end{equation}
которые есть ``квантования'' вида (\ref{QHamst}) с  $\varepsilon=1$, определяемые  гамильтонианами (\ref{Kamone}) и (\ref{Kamtwo}) ---   
символически  (\ref{Kamz}) и (\ref{Kamt}) можно записать в виде уравнений 
$$\Gamma_{t_1}=K_1(t_1,t_2,X,Y,-\frac{\partial}{\partial X},-\frac{\partial}{\partial Y})\Gamma, \quad \Gamma_{t_2}=K_2(t_1,t_2,X,Y,-\frac{\partial}{\partial X},-\frac{\partial}{\partial Y})\Gamma.$$

С помощью формулы
$$\Gamma=\exp{(-\frac{X^4+Y^4}{4}-\frac{t_2^2+t_2(X^2+Y^2)+t_1(X+Y)}{2}-\frac{t_1^2t_2
}{4})}W$$
``квантования'' (\ref{Kamz}) и (\ref{Kamt}) сводятся к совместной паре  уравнений
$$2(X-Y)W_{t_1}=W_{XX}-W_{YY}-[X^6-Y^6+2t_2(X^4-Y^4)+$$
$$+t_1(X^3-Y^3)+(t_2^2+2\nu))(X^2-Y^2)]W,$$
$$2(X-Y)W_{t_2}=XW_{YY}-YW_{XX}+W_X-W_Y-[X(Y^6+2t_2Y^4+t_1Y^3+$$
$$+(t_2^2+2\nu)Y^2-\frac{t_1^2}{4})-Y(X^6+2t_2X^4+t_1X^3+
(t_2^2+2\nu)X^2-\frac{t_1^2}{4})]W.$$
После перехода к независимым переменным (\ref{channtwo}), замен 
$$X=-(4\beta)^{1/4}\exp(i\pi/8)\zeta,\quad Y=-(4\beta)^{1/4}\exp(i\pi/8)\eta$$ 
и $c=-8i\beta\nu$ 
последняя пара  сводится к паре  (\ref{KNez}), (\ref{KNet}).

\mbox{З а м е ч а н и е 2.} Скорее всего, аналогичным образом может быть описана связь между парой уравнений (\ref{KGPez}), (\ref{KGPet}) и  ``квантованиями'', определяемыми совместной парой $H(9/2)$ гамильтоновых систем статьи \cite{kimu}.

\section{Заключительные замечания}

{\bf 3.1.}  При $\varepsilon=1$ ``квантования'' (\ref{murze}) из \cite{Difufa} получаются не только из уравнений  (\ref{Schroo}), но и  из уравнений $\varepsilon\Psi_z =
H(z,x,\varepsilon\frac{\partial}{\partial x})\Psi,$
для которых при произвольных значениях 
 $\varepsilon$  и определенных редукциях уравнений Пенлеве в \cite{Nag} были построены серии явных решений. В связи с этим результатом \cite{Nag}, а также в связи с  разделом 2 статьи \cite{Ufmj} и результатами статьи \cite{kwin} подчеркнем следующее обстоятельство:
 
при $\varepsilon=5/54$ (\ref{qnGpt}) можно переписать как уравнение  
$$\varepsilon \Psi_{t}=\varepsilon^2[
-\frac{\Psi_{xx}}{2}-\frac{1}{2}\frac{\partial}{\partial x}(x \Psi_{y})
-\frac{1}{2}x\Psi_{xy}+(y-\frac{5t}{36})\Psi_{yy}]
+\varepsilon\frac{5\Psi_y}{108}+$$
$$+[\frac{x^5}{2}-2xy^2
-\frac{5tx^3}{36}+\frac{5txy}{18}-\frac{5zx^2}{216}-\frac{5zy}{108}]\Psi.$$

Это значит, что наряду с представлениями в виде ``квантований'' (\ref{QHamst}) пары (\ref{qnGpz}), (\ref{qnGpt})  можно символически записать и как уравнения:  
$$\varepsilon\Psi _z=H_1(z,t,x,y,\varepsilon \frac{\partial}{\partial x},\varepsilon  \frac{\partial}{\partial y})\Psi,\quad \varepsilon\Psi _t=H_2(z,t,x,y,\varepsilon\frac{\partial}{\partial x},\varepsilon \frac{\partial}{\partial y})\Psi.$$ 

Возможно, подобную символическую запись допускают и какие-нибудь ``квантования'',   определяемые гамильтонианами, эквивалентыми  гамильтонианам (\ref{Kamone}) и (\ref{Kamtwo}).

{\bf 3.2.} При описании выше ``квантований'' двух высших  аналогов уравнений Пенлеве ключевой явилась замена (\ref{NovDP}), которая ранее в близких ситуациях для несколько иных целей использовалась Д.
П. Новиковым в \cite{Novd} (cм. также формулу (2.3.36) работы \cite{sato}).  Не факт, что лишь с помощью подобной замены также легко прояснится вопрос о справедливости  ``квантований'' вида (\ref{QHamst}) для всех  высших гамильтоновых изомонодромных аналогов Пенлеве с двумя степенями свободы, которым соответствуют уравнения ИДМ в матрицах размером $2\times2$ ( в частности, для всех, рассматривавшихся в \cite{kimu}).

Пока не видно, как результаты данной работы могут быть обобщены на случаи изомонодромных гамильтоновых ОДУ с числом степеней свободы, больших  2, пусть даже для ситуаций, отвечающих уравнениям ИДМ в матрицах размером $2\times2$
(например, для представителей иерархий $P_1^n$, $P_2^n$ c $n>2$). В связи с только что сказанным отметим, что  в случае систем Шлезингера \cite{schle}, соответствующих уравнениям ИДМ в матрицах размером $2\times2$, эти линейные уравнения ИДМ задают \cite{Novd} решения уравнений типа Белавина --- Полякова --- Замолодчикова \cite{bel} минимальной конформной теории поля с центральным зарядом алгебры Вирасоро, равным единице. Но не связаны ли данные уравнения ИДМ также и с какими-либо ``квантованиями'', определяемыми гамильтоновыми структурами  соответствующих систем Шлезингера? 

Совсем неисследованным остается и вопрос о ``квантованиях'' изомонодромных гамильтоновых ОДУ, отвечающих уравнениям ИДМ в матрицах размером  $m\times m$ при $m>2.$

\end{document}